\documentclass{iaus}
\usepackage{graphicx}
\usepackage{natbib}

\title[Total and spectral solar irradiance variations] 
{Mechanisms for total and spectral solar irradiance variations}

\author[Margit Haberreiter]  
{Margit Haberreiter}
\affiliation{Laboratory for Atmospheric and Space Physics, University of Colorado, 1234 Innovation Drive, Boulder, CO, 80303, USA \\ email: {\tt haberreiter@lasp.colorado.edu}} 

\pubyear{2009}
\volume{264}  
\pagerange{???--???}
\jname{Solar and Stellar Variability – Impact on Earth and Planets}
\editors{A.C. Editor, B.D. Editor \& C.E. Editor, eds.}
\begin{document}

\maketitle

\begin{abstract}
The total and spectral irradiance varies over short time scales, i.e. from days to months, and longer time scales from years to decades, centuries, and beyond. In this talk we review the current understanding of irradiance changes from days to decades. We present the current status of observations and discuss proposed reconstruction approaches to understand these variations. The main question that ultimately needs to be answered is what are the physical processes that could explain the enhanced heating of the photosphere, chromosphere, transition region, and corona, leading to a change in the solar radiative output at various wavelengths. As semi-empirical models allow us to reproduce the solar spectrum over a broad wavelength range, they offer a powerful tool to determine the energy necessary to heat certain layers and at the same time balance the radiative losses. 
\end{abstract}

\firstsection 
\section{Introduction}
The incoming solar radiation is the main energy driving the Earth's climate system. This radiation varies as a function of wavelengths on timescales of days to months, and years to decades and centuries. While the variations of the total solar irradiance (TSI) cannot be the reason for the increase of the mean global temperature over the past 20 years \citep{LockwoodFroehlich2007}, the variations of the solar spectral irradiance (SSI) are considered to have an effect on the Earth's climate system. For example, \cite{Egorova2004}, and \cite{Austin2008} demonstrate that the Earth's atmosphere shows an effect on the solar cycle variability in the UV spectral range. The effect of a long-term trend of the solar spectral variability on the Earth's climate system are however not yet fully understood. This is partly due to the lack of a sufficiently long and precise time series of the SSI that is needed as input for Global Circulation Models (GCMs). Therefore, reconstructions of the TSI and SSI for time scales of decades and longer are essential for fully understanding of the Sun's role in the Earth's climate system. 

Another important aspect of the SSI is the varying EUV radiation that influences the thermosphere and ionosphere \citep{FullerRowell2004}. Due to the variations of the incoming EUV radiation, the neutral density in the Earth's upper atmosphere changes. As the motion of satellites depends on the neutral density, being able to nowcast and forcast the density allows a more precise localization of the satellites. This clearly improves the reliablity of satellite operation.

Finally, understanding the effects of the varying SSI on the Earth's atmosphere and climate system has received increasing interest not only for the Earth, but also with respect to the conditions of the atmospheres of other planets. 

In the next section the latest observations are briefly mentioned. In Sec.\,\ref{sec:mech} the current status of the reconstruction models for the TSI and SSI is presented. Then, in Sec.\,\ref{sec:euv} we present latest calculations for the EUV spectral range. Finally, in Sec.\,\ref{sec:disc} we discuss the latest reconstruction approaches in a broader perspective.
\section{Observations}
\subsection{Total Solar Irradiance Variations}
\begin{figure}[t!]
\begin{center}
 \includegraphics[width=0.8\textwidth]{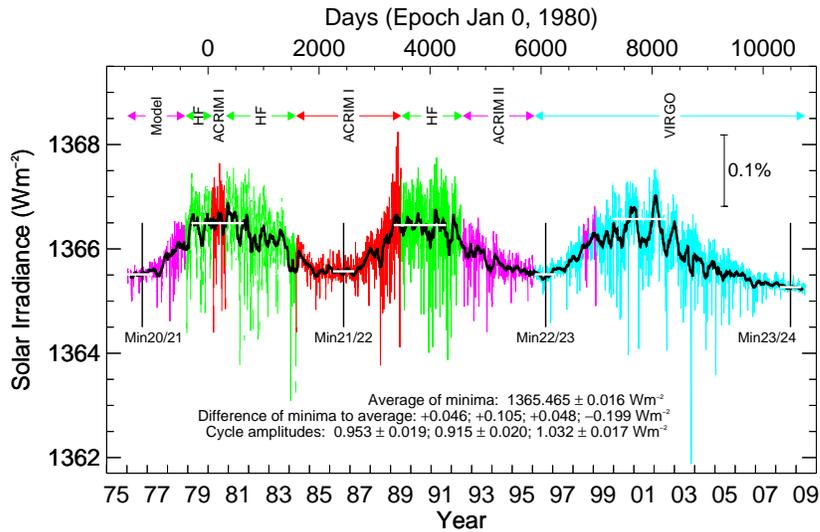} 
 \caption{Latest available update of the TSI composite since the start of its measurements from space. The colored lines indicate daily measurements taken by different instruments the composite is based on; for details see \cite{Froehlich2009}.}
   \label{fig1}
\end{center}
\end{figure}

The TSI is the wavelength integrated radiation emitted by the Sun measured at a distance of 1~AU. For more than three decades it has been measured from space. The measurements were taken by a number of different instruments; for details see e.g. \cite{Domingo2009}, Sec.~2.1. Fig.\,\ref{fig1} shows the composite compiled by \cite{Froehlich2009} showing a clear 11-year solar cycle variability. This variability is considered to be a result of the changing contribution of surface features, such as the quiet and active network, sunspots, plage, and faculae, to the overall emergent intensity.

\subsection{Spectral Solar Irradiance Variations}
The variability of the SSI is a strong function of wavelength. Therefore, in order to study the response of the Earth's atmosphere over the solar cycle the SSI has to be continuously observed over decades with sufficient spectral resolution and over a broad wavelength range. For a detailed list of past and future missions dedicated to observing the SSI see \cite{Domingo2009}, Sec.~2.3. A few of the most recent instruments are listed below. In the UV, a continuous time series from space was only started in 1991 with the launch of the Upper Atmosphere Research Satellite \cite[UARS]{Willson1994}, having the Solar Ultraviolet Spectra Irradiance Monitor \citep[SUSIM]{Brueckner1993} and the Solar Stellar Irradiance Comparison Experiment \citep[SOLSTICE]{Rottman1993} onboard. With the launch of the SOlar Radiation and Climate Experiment \cite[SORCE]{Rottman2005} the spectral range of the SSI from 1~nm to 2,700~nm was for the first time covered for continuous observations. The Solar Auto-Calibrating EUV/UV Spectrometer (SolACES) and SOLar SPECtrum (SOLSPEC) experiments on SOLAR/ISS \citep{Thuillier2003, Schmidtke2006} that was launched in February 2009 measure the solar spectrum from 16 to 226~nm (in 4 separate channels) and 180 to 3000~nm. 

Recently, \cite{Harder2009} discovered from the SORCE/SIM \citep{Harder2005SoPh} observations that different wavelength intervals of the solar spectrum show opposite trends in the solar cycle variability. These new findings show that the TSI variability is the sum of a SSI variability, which partly cancels due to its opposite sign. These findings pose a challenge for recontruction models that need to be able to reproduce this effect.

The continuous time series of SSI provided by SORCE is extremely valuable for the study of the solar radiation effects on the Earth's climate. For climate studies the ultimate goal is to reconstruct solar irradiance as far back in time as possible to study its long-term effects on the Earth's climate. Thus, precise reconstructions of the SSI over decades and beyond are necessary. In the next section the available reconstruction models are discussed.
\section{Models for irradiance variations}\label{sec:mech}
\subsection{Total Solar Irradiance Variations}\label{sec:tsi}
With the availability of precise TSI and SSI measurements from space since the 1970s, modeling these variations became a challenging task. The early models were based on indices such as e.g. the sunspot number (SSN), the Mg~II core-to-wing ratio, and the solar F10.7~cm radio flux. These indices are associated with solar activity, but not directly with the emitted radiation. \cite{FroehlichLean2004} could explain the TSI variations by a combination of the sunspot darkening derived from the SSN and the facular brightening based on a long-term and short-term analysis of the Mg~II index. 
The SATIRE model \citep[and references therein]{Solanki2005} uses intensity spectra calculated with the ATLAS9 code by \cite{Kurucz1991} to describe the intensity emitted by different solar surface features, i.e. quiet Sun, sunspot umbra and penumbra, network, and faculae. SOHO/MDI and Kitt Peak magnetograms are employed to derive the surface area covered by the features as a function of position on the disk. For the calculation of the intensity spectra, local thermodynamic equilibrium (LTE) is assumed.
These models are very successful and can explain the TSI variability almost entirely due to the changes of the solar surface magnetic field. 

Evidence that the Sun varies on longer time scales is seen in radionuclide data, e.g. $^{10}$Be from ice cores \citep{Beer2000}. Radionuclides are produced when cosmic ray particles enter the Earth's atmosphere and interact with the atoms in the atmosphere, thereby also producing neutrons. The flux of the cosmic ray particles is controlled by the heliospheric magnetic field (HMF) and the geomagnetic field. In fact, the production rate of the radionuclides is anti-correlated with the extent of the HMF. It has been shown by \cite{Beer2000} that the rates of neutrons and $^{10}$Be behave very similar. This fact allowed \cite{Schoell2007} to reconstruct the variability of the TSI based on neutron monitor data and the SSN. Based on $^{10}$Be data, recently \cite{Steinhilber2009} presented a TSI reconstruction for the past 9300 years.    

\subsection{Spectral Solar Irradiance Variations}\label{sec:rec_proxy}
\subsubsection{Reconstruction from magnetograms}
\begin{figure}[t!]
\begin{center}
 \includegraphics[width=0.8\textwidth]{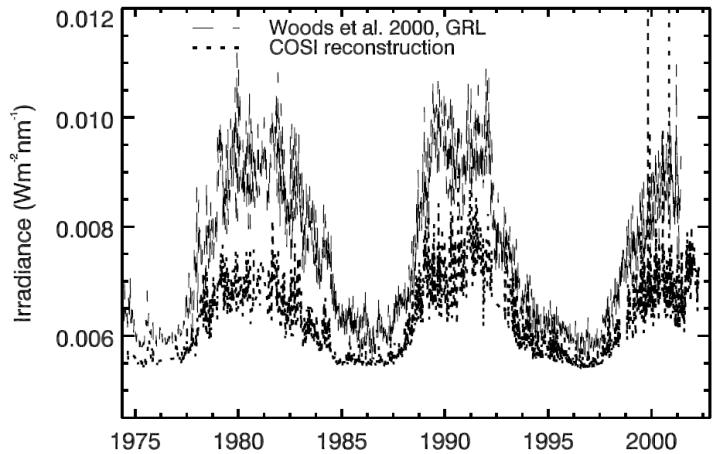} 
 \caption{Reconstruction of Lyman\,$\alpha$ based on NLTE spectral synthesis with the COSI code and the feature identification from magnetograms as employed in the SATIRE model. Adopted from \cite{Hab2005AdSpR}.}
   \label{fig:lya}
\end{center}
\end{figure}
The calculation of solar intensity spectra based on the assumption of LTE, as described in Sec.\,\ref{sec:tsi}, is not suitable for the UV and shorter wavelengths. Moreover, the formation of negative hydrogen, the main continuum opacity source in the visible and IR, depends on the electron density, which in turn depends on the ionizing UV radiation. As the latter shows non-LTE (NLTE) effects, so does the formation of negative hydrogen; for details see e.g. \cite{Shapiro2009}. Thus, accounting for the NLTE effects is essential for the synthesis of the correct solar spectrum, from the UV to the IR. These effects clearly also influence the solar cycle variation of the SSI.
 
Nevertheless, within the SATIRE model a method has been developed that allows extrapolating the SSI for wavelengths down to 115~nm. Based on this method, \cite{Krivova2009} present the reconstruction of the TSI and SSI for the years from 1974 to 2007. Again, the SATIRE model is very successful in describing the variability of the TSI and SSI. However, one has to keep in mind that the physical processes involved in the irradiance changes are very complex, and, in order to understand them they need to be accounted for in the models in full detail. 
 
In order to improve the spectral synthesis \cite{Hab2005AdSpR} extended the work by \cite{Krivova2003} and calculated solar spectra with the NLTE radiative transfer COde for Solar Irradiance \citep[COSI]{Haberreiter2008a,Haberreiter2008b} using the atmosphere structures by \cite{Fontenla1999}. Based on the magnetogram analysis used in SATIRE they reconstructed the SSI for the wavelength range from 100 to 400\,nm. Fig.\,\ref{fig:lya} shows their reconstruction for Lyman $\alpha$ and compares it to the composite by \cite{Woods2000}. Following the SATIRE approach, the reconstruction is based on the changing intensity contributions for the quiet Sun, sunspots, and faculae. From Fig.\,\ref{fig:lya} it is clear that the reconstruction underestimates the solar cycle variability by up to a factor of 2. This reconstruction however did not include the quiet and active network contribution. Whether the additional quiet and active network contribution is sufficient to explain the observed solar cycle variability still remains open. Another important point is that all reconstruction approaches discussed above assume a constant quiet Sun intensity over the solar cycle. However, work by \cite{Schuehle2000} indicates that the quiet Sun radiance changes over the solar cycle. If this is confirmed the reconstruction approaches certainly need to be revised.

\subsubsection{Reconstruction from intensity images}\label{sec:rec_intens}
The analysis of magnetograms is only an indirect way to determine the brightness components responsible for irradiance variations. In fact the direct measurement of the intensity of all of the components allows a better understanding of their nature. 

The Presicion Solar Photometric Telescope \cite[PSPT]{PSPT,Ermolli2003,White2000SSRv} at Rome and M.~Loa Observatories take images in a narrow-band Ca~II filter at 393~nm and a continuum wavelengths at 607~nm. These measurements allow the identification of solar surface features based on the intensity measured in each pixel. For example, the Solar Radiation Physical Modeling (SRPM) system provides tools to identify the surface features, such as intergranular cell, quiet and active network, faculae, plage, sunspot umbra and penumbra, which are identified through their contrast as a function of position on the disk \citep{Fontenla2005,Fontenla2009b,Harder2005}. 

\begin{figure}[t!]
\begin{center}
 \includegraphics[width=0.8\textwidth]{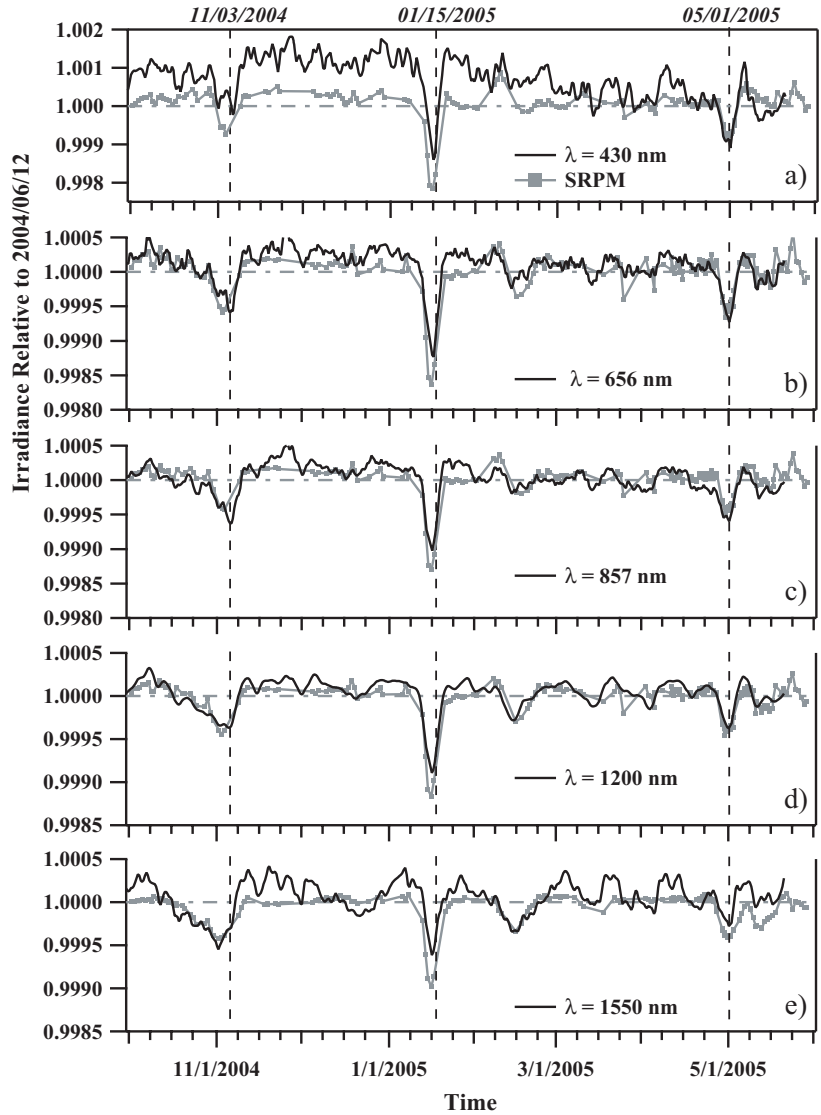} 
 \caption{Comparison of the SRPM reconstruction based on PSPT images and SIM/SORCE spectral irradiance measurements for different wavelengths \citep{Harder2005}. The relative changes of the SSI are in solid agreement with the observation, indicating that there is a firm understanding of the SSI variability at these wavelengths.}
   \label{fig:harder}
\end{center}
\end{figure}

The latest set of 1D atmosphere structures \citep{Fontenla2009b} facilitates calculating the observed intensity and center-to-limb behavior for each of the surface features as identified from the PSPT observations. Please note that the latest set of models does not represent the quiet Sun with one model. In fact the intensity spectrum for the quiet Sun is considered to be a combination of intergranular cell (Model B), quiet network (Model D) and active network (Model F). 
Fig.\,\ref{fig:harder} shows the reconstruction based on this scheme for different wavelengths with an earlier set of atmosphere structures. The agreement is very robust, with slightly larger deviations for 430~nm. The approches discussed so far are used to reconstruct the TSI and SSI for past times, but there is also an increased intereset in nowcasting and forecasting the solar irradiance, in particular for shorter wavelengths. 

Recently, a lot of progress has been made in forecasting active regions on the far side of the Sun \citep{LindseyBrown2000}. Along with the observations of backscattered Lyman\,$\alpha$ \citep{Quemerais2002}, recently \cite{Fontenla2009a} developed a tool that is capable of forecasting the Lyman\,$\alpha$ radiation for a solar rotation. The great advantage of this method is that it accounts for growing and decaying active regions that have not yet appeared on the near side. This opens the possibility to forecasting the full solar spectrum in the near future. One important application for forecasting the EUV is to be able to determine the neutral density in the ionosphere, which in turn permits a more precise satellite positioning. 
\subsection{SSI variations in the EUV}\label{sec:euv}
\begin{figure}[!t]
\begin{center}
 \includegraphics[width=1.\textwidth]{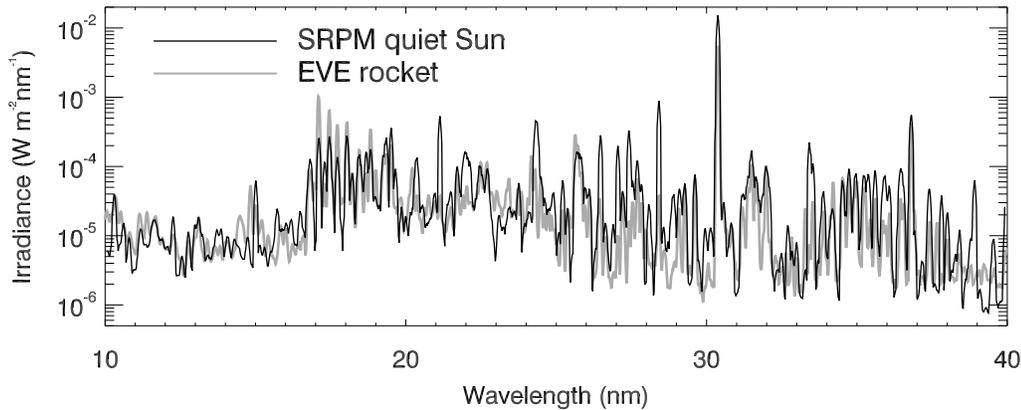} 
 \caption{Comparison of the synthetic spectrum (thin black line) with the spectrum observed with the EVE instrument during a rocket calibration flight on April 14, 2008 (thick gray line). Adopted from \cite{HabFont2009}.}
   \label{fig:euv}
\end{center}
\end{figure}

Similar to the reconstruction approach presented in Sec.\,\ref{sec:rec_intens}, the irradiance in the EUV is considered to be the sum of all intensities emitted by the different features at and beyond the solar disk. Fig.\,\ref{fig:euv} shows the synthetic spectrum calculated with the SRPM system in spherical symmetry compared to the observed spectrum measured with the EVE rocket instrument during a calibration flight on April 14, 2008 \citep{Chamberlin2009}. For details on the spectral synthesis see \cite{HabFont2009}. We are currently in the process of calculating intensity spectra for various types of coronal features. Based on the feature identification from EIT images, this will finally allow us to calculate the irradiance in the EUV. 

\begin{figure}[t!]
\begin{center}
 \includegraphics[height=0.5\textwidth]{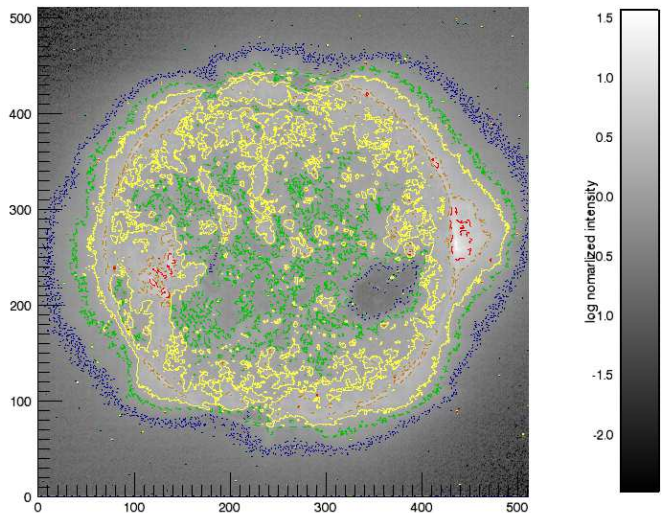} 
 \caption{Coronal features identified from a SOHO/EIT image take at 17.1~nm. The contour levels correspond to -0.3 (blue dotted line, coronal hole), 0 (green dashed line, quiet corona), 0.2 (yellow solid line, coronal network), 0.5 (orange dashed-dotted line, hot corona) and 0.75 (red long dashes, super-hot corona) of the logarithm of the intensity normalized to the median. These contour levels provide a good first measure to identify different coronal features.}
   \label{fig:eit_contours}
\end{center}
\end{figure}

Depending on the wavelength, the EUV is emitted by different layers, i.e. the chromosphere, transition region, or corona. SOHO/EIT \citep{EIT} observes at four wavelengths, i.e. 17.1~nm, 19.5~nm, 28.4~nm, and 30.4~nm. The wavelength range at 30.4~nm represents the cooler chromosphere, while 28.4~nm represents the hotter corona. It is clear that the feature identification in these regimes has to be different than in the case of the PSPT images. Nevertheless, the intensity distribution in each image is a powerful measure to define thresholds that are characteristic for the chromospheric and coronal features. Fig.\,\ref{fig:eit_contours} shows the possibility to identify coronal features from a SOHO/EIT image observed on July 30, 2007 at 17.1~nm. The gray-scale is the logarithm of the intensity normalized to its median. The colored contour lines are thresholds that define coronal features such as coronal and equatorial holes (blue), quiet corona (green), coronal network (yellow), and hot and super-hot corona (orange and red). As the area covered by these features changes over the solar cycle, so does the disk-integrated radiation emitted by the Sun. In the EUV A difficulty arises from the fact that with increasing distance the extended corona gets increasingly optically thin, which makes a feature identification more problematic. Furthermore, it is clear that a spherical line-of-sight integration that takes into account the extended corona is very important. 
\section{Discussion}\label{sec:disc}
Comprehending the irradiance variations ultimately depends on understanding the heating mechanisms that are responsible for an enhanced brightness of the different features on the solar disk. It is understood that the reason why sunspots are dark is due to the suppression of convective heat flow in the presence of strong magnetic field. On the other hand quiet and active network, plage, and faculae show an increased contrast. All these features are associated with enhanced magnetic field. 

However, the contrast of magnetic elements is a function of wavelength. Sunspots show a negative contrast in the visible wavelength ranges but are typically bright in the UV, in particular in Lyman~$\alpha$. The fact that the contrast of solar surface features changes as a function of wavelength indicates that different heating processes are involved in different layers of the solar atmosphere. 

The 1D atmosphere structures by \cite{Fontenla2009b} are suitable to reproduce solar spectra at a moderate spacial resolution of 2'' from the FUV to the IR. These structures are time-independent representations of an average solar atmosphere, and as such they cannot account for any of the possibly dynamic processes involved in the chromospheric and coronal heating. The additional strength of these models is, however, that they might serve as a diagnostic tool to determine how much heating needs to be provided by reasonable mechanisms in order to sustain certain temperatures in different layers of the solar atmosphere and at the same time balance the radiative losses. 

Some of the proposed heating mechanisms are low-frequency acoustic waves \citep{Jefferies2006}, the Farley-Buneman instability \citep{Fontenla2008}, coronal heating from Alfv\'en waves \cite{Cranmer2007}, nanoflares \cite{PatsourakosKlimchuk2009}, and heating processes of the corona that are rooted in the chromosphere \citep{DePontieu2009}. Ultimately, to fully understand the physics of the solar atmosphere, the physical processes of various heating mechanisms need to be included in the forward modeling, which eventually will allow us to reproduce solar spectra over a broad wavelength range. Additionally, the heating mechanisms also have to explain the solar cycle variability. Here, the physical properties of the 1D atmosphere structures offer good diagnostics about the temperature and pressure in certain regimes of the solar atmosphere. On the other hand, forward modeling using 3D MHD codes allows accounting for dynamic heating processes. Combining the insight that has been gained from both approaches has certainly great potential to advance our understanding of the physics of the solar atmosphere, and ultimately the variability of the TSI and SSI. 
\section{Conclusions}\label{sec:conc}
Our understanding of the TSI and SSI variations over shorter and longer timescales has advanced substantially over the past decades. Starting with solar activity proxies to reproduce these variations, we are now able to use detailed radiative transfer calculations to account for the irradiance variations. One of the remaining open questions is whether the quiet Sun radiance shows a long-term trend. If this is confirmed, then the current reconstruction models need to be refined to account for this effect. 

The increased precision of the TSI and SSI measurements builds the base for the understanding of the processes that are behind its variability. The upcoming irradiance measurements by Picard/PREMOS \citep{Schmutz2009}, PROBA2/LYRA \citep{Hochedez2006}, and SDO/EVE \citep{Woods2006} will certainly improve our understanding of the TSI and SSI variability. SDO/AIA will provide images with an unprecedented combination of spatial and temporal resolution, and thus be very valuable for advancing our understanding of the physics behind the irradiance variations. 
\begin{acknowledgments}
MH gratefully acknowledges travel support by the International Astronomical Union. MH also thanks the Solar Influences Group at LASP for helpful discussions.
\end{acknowledgments}


\begin{thebibliography}{}

\bibitem[Austin \etal\ (2008)]{Austin2008}
{Austin, J., Tourpali, K., Rozanov, E., Akiyoshi, H., Bekki, S., Bodeker, G., Br{\"u}hl, C., Butchart, N., Chipperfield, M., Deushi, M., Fomichev, V.I., Giorgetta, M.A., Gray, L., Kodera, K., Lott, F., Manzini, E., Marsh, D., Matthes, K., Nagashima, T., Shibata, K., Stolarski, R.S., Struthers, H., \& Tian, W.} 2008, \textit{Journal of Geophysical Research (Atmospheres)}, 113, 11306

\bibitem[Beer (2000)]{Beer2000}
{Beer, J.} 2000, \textit{Space Science Reviews}, 94, 53

\bibitem[Brueckner \etal\ (1993)]{Brueckner1993}
{Brueckner, G.E., Edlow, K.L., Floyd, L.E., Lean, J.L., \& Vanhoosier, M.E.} 1993, \textit{Journal of Geophysical Research}, 98, 10695

\bibitem[Chamberlin \etal\ (2009)]{Chamberlin2009}
{Chamberlin, P.C., Woods, T.N., Crotser, D.A., Eparvier, F.G., Hock, R.A., Woodraska, D.L.} 2009, \textit{Geophysical Research Letters}, 36, 5102.

\bibitem[Coulter \etal\ (1996)]{PSPT}
{Coulter, R.L., Kuhn, J.R., \& Lin, H.} 1996, \textit{Bulletin of the American Astronomical Society}, 28, 912

\bibitem[Cranmer \etal\ (2007)]{Cranmer2007}
{Cranmer, S.R., van Ballegooijen, A.A., \& Edgar, R.J.} 2007, \textit{Astrophysical Journal Supplement Series}, 171, 520

\bibitem[Delaboudini{\`e}re \etal\ (1995)]{EIT}
{Delaboudini{\`e}re, J.P., Artzner, G.E., Brunaud, J., Gabriel, A.H., Hochedez, J.F., Millier, F., Song, X.Y., Au, B., Dere, K.P., Howard, R.A., Kreplin, R., Michels, D.J., Moses, J.D., Defise, J.M., Jamar, C., Rochus, P., Chauvineau, J.P., Marioge, J.P., Catura, R.C., Lemen, J.R., Shing, L., Stern, R.A., Gurman, J.B., Neupert, W.M., Maucherat, A., Clette, F., Cugnon, P., van Dessel, E.L.} 1995, \textit{Solar Physics}, 162, 291

\bibitem[De Pontieu \etal\ (2009)]{DePontieu2009}
{De Pontieu, B., McIntosh, S.W., Hansteen, V.H., \& Schrijver, C.J.} 2009, \textit{Astrophysical Journal}, 701, L1.

\bibitem[Domingo \etal\ (2009)]{Domingo2009}
{Domingo, V., Ermolli, I., Fox, P., Fr{\"o}hlich, C., Haberreiter, M., Krivova, N., Kopp, G., Schmutz, W., Solanki, S.K., Spruit, H.C., Unruh, Y., \& V{\"o}gler, A.} 2009, \textit{Space Science Reviews}, 145, 337

\bibitem[Egorova \etal\ (2004)]{Egorova2004}
{Egorova, T., Rozanov, E., Manzini, E., Haberreiter, M., Schmutz, W., Zubov, V., \& Peter, T.} 2004, \textit{Geophysical Research Letters}, 31, 6119

\bibitem[Ermolli \etal\ (2003)]{Ermolli2003}
{Ermolli, I., Berrilli, F., \& Florio, A.} 2003, \textit{Astronomy and Astrophysics}, 412, 857


\bibitem[Fontenla \etal\ (1999)]{Fontenla1999}
{Fontenla, J., White, O.R., Fox, P.A., Avrett, E.H., \& Kurucz, R.L.} 1999 \textit{Astrophysical Journal}, 518, 480

\bibitem[Fontenla \etal\ (2005)]{Fontenla2005}
{Fontenla, J. \&  Harder, G.} 2005, \textit{Memorie della Societa Astronomica Italiana}, 76, 826

\bibitem[Fontenla \etal\ (2008)]{Fontenla2008}
{Fontenla, J.M., Peterson, W.K., \& Harder, J.} 2008, \textit{Astronomy and Astrophysics}, 480, 839

\bibitem[Fontenla \etal\ (2009a)]{Fontenla2009a}
{Fontenla, J.M., Qu{\'e}merais, E., Gonz{\'a}lez Hern{\'a}ndez, I.,	Lindsey, C., \& Haberreiter, M.} 2009a, \textit{Advances in Space Research}, 44, 457

\bibitem[Fontenla \etal\ (2009b)]{Fontenla2009b}
{Fontenla, J.M., Curdt, W., Haberreiter, M., Harder, J., \& Tian, H.} 2009b, \textit{Astrophysical Journal}, accepted for publication

\bibitem[Fr{\"o}hlich \& Lean (2004)]{FroehlichLean2004}
{Fr{\" o}hlich, C. \& Lean, J.} 2004, \textit{The Astronomy and Astrophysics Review}, 12, 273

\bibitem[Fr{\"o}hlich (2009)]{Froehlich2009}
{Fr{\"o}hlich, C.} 2009, \textit{Astronomy and Astrophysics}, 501, L27

\bibitem[Fuller-Rowell \etal\ (2004)]{FullerRowell2004}
{Fuller-Rowell, T., Solomon, S., Roble, R., \& Viereck, R.} 2004, in: {J.M. Pap, P. Fox, C. Fr{\"o}hlich, H.S. Hudson, J. Kuhn, J. McCormack, G. North, W. Sprigg, \& S.T. Wu} (eds.), \textit{Solar Variability and its Effects on Climate. Geophysical Monograph 141}, p.~141

\bibitem[Haberreiter \etal\ (2005)]{Hab2005AdSpR}
{Haberreiter, M., Krivova, N.A., Schmutz, W., \& Wenzler, T.} 2005, \textit{Advances in Space Research}, 35, 365

\bibitem[Haberreiter \etal\ (2008a)]{Haberreiter2008a}
{Haberreiter, M., Schmutz, W., \& Hubeny, I.} 2008a, \textit{Astronomy and Astrophysics}, 492, 833

\bibitem[Haberreiter \etal\ (2008b)]{Haberreiter2008b}
{Haberreiter, M., Schmutz, W., \& Kosovichev, A.G.} 2008b, \textit{Astrophysical Journal}, 675, L53

\bibitem[Haberreiter \& Fontenla (2009)]{HabFont2009}
{Haberreiter, M. \& Fontenla, J.} 2009, in: I. Hubeny, J.M. Stone, K. MacGregor, \& K. Werner (eds.), \textit{Recent Directions in Astrophysical Quantitative Spectroscopy and Radiation Hydrodynamics}, p.\,355

\bibitem[Harder \etal\ (2005a)]{Harder2005}
{Harder, J., Fontenla, J., White, O., Rottman, G., \& Woods, T.} 2005a, \textit{Memorie della Societa Astronomica Italiana}, 76, 735

\bibitem[Harder \etal\ (2005b)]{Harder2005SoPh}
{Harder, J., Lawrence, G., Fontenla, J., Rottman, G., \& Woods, T.} 2005b, \textit{Solar Physics}, 230, 141

\bibitem[Harder \etal\ (2009)]{Harder2009}
{Harder, J.W., Fontenla, J.M., Pilewskie, P., Richard, E.C., \& Woods, T.N.} 2009, \textit{Geophysical Research Letters}, 36, 7801

\bibitem[Hochedez \etal\ (2006)]{Hochedez2006}
{Hochedez, J.-F., Schmutz, W., Stockman, Y., {Sch{\"u}hle}, U., Benmoussa, A., Koller, S., Haenen, K., Berghmans, D., Defise, J.-M., Halain, {J.-P.}, Theissen, A., Delouille, V., Slemzin, V., Gillotay, D., Fussen, D., Dominique, M., Vanhellemont, F., McMullin, D., Kretzschmar, M., Mitrofanov, A., Nicula, B., Wauters, L., Roth, H., Rozanov, E., R{\"u}edi, I., Wehrli, C., Soltani, A., Amano, H., van der Linden, R., Zhukov, A., Clette, F., Koizumi, S., Mortet, V., Remes, Z., Petersen, R., Nesl{\'a}dek, M., D'Olieslaeger, M., Roggen, J., Rochus, P.} 2006, \textit{Advances in Space Research}, 37, 303

\bibitem[Jeffe\-ries \etal\ (2006)]{Jefferies2006}
{Jefferies, S.M., McIntosh, S.W., Armstrong, J.D., Bogdan, T.J., Cacciani, A., Fleck, B.} 2006, \textit{Astrophysical Journal}, 648, L151

\bibitem[Krivova \etal\ (2003)]{Krivova2003}
{Krivova, N.A., Solanki, S.~K., Fligge, M., \& Unruh, Y.C.} 2003, \textit{Astronomy and Astrophysics}, 399, L1

\bibitem[Krivova \etal\ (2009)]{Krivova2009}
{Krivova, N.A., Solanki, S.K., Wenzler, T., \& Podlipnik, B.} 2009, \textit{Journal of Geophysical Research (Atmospheres)}, 114, 1

\bibitem[Kurucz \etal\ (1991)]{Kurucz1991}
{Kurucz, R.L.} 1991, in: \textit{NATO ASIC Proc. 341: Stellar Atmospheres - Beyond Classical Models}, p.~441

\bibitem[Lindsey \& Brown (2000)]{LindseyBrown2000}
{Lindsey, C. \& Braun, D.C.} 2000, \textit{Science}, 287, 1799

\bibitem[Lockwood \& Fr{\"o}hlich (2007)]{LockwoodFroehlich2007}
{Lockwood, M. \& Fr{\" o}hlich, C.} 2007, \textit{Proceedings of the Royal Society A}, 463, 2447

\bibitem[Patsourakos \& Klimchuk (2009)]{PatsourakosKlimchuk2009}
{Patsourakos, S. \& Klimchuk, J.A.} 2009, \textit{Astrophysical Journal}, 696, 760

\bibitem[Qu{\'e}merais \& Bertaux (2002)]{Quemerais2002}
{Qu{\'e}merais, E., \& Bertaux, J.L.} 2002, \textit{Geophysical Research Letters}, 29, 2

\bibitem[Rottman \etal\ (1993)]{Rottman1993}
{Rottman, G.J., Woods, T.N., \& Sparn, T.P.} 1993, \textit{Journal of Geophysical Research}, 98, 10667

\bibitem[Rottman (2005)]{Rottman2005}
{Rottman, G.} 2005, \textit{Solar Physics}, 230, 7

\bibitem[Rozanov \etal\ (2006)]{Rozanov2006}
{Rozanov, E., Egorova, T., Schmutz, W., \& Peter, T.} 2006, \textit{Journal of Atmospheric and Solar-Terrestrial Physics}, 68, 2203


\bibitem[Schmidtke \etal\ (2006)]{Schmidtke2006}
{Schmidtke, G., Fr{\"o}hlich, C., \& Thuillier, G.} 2006, \textit{Advances in Space Research}, 37, 255

\bibitem[Schmutz \etal\ (2009)]{Schmutz2009}
{Schmutz, W., Fehlmann, A., H{\"u}lsen, G., Meindl, P., Winkler, R., Thuillier, G., Blattner, P., Buisson, F., Egorova, T., Finsterle, W., Fox, N., Gr{\"o}bner, J., Hochedez, J.-F., Koller, S., Meftah, M., Meisonnier, M., Nyeki, S., Pfiffner, D. Roth, H., Rozanov, E., Spescha, M., Wehrli, C., Werner, L., \& Wyss, J.U.} 2009, \textit{Metrologia}, 46, 202

\bibitem[Sch\"oll \etal\ (2007)]{Schoell2007}
{Sch{\"o}ll, M., Steinhilber, F., Beer, J., Haberreiter, M., \& Schmutz, W.} 2007, \textit{Advances in Space Research}, 40, 996

\bibitem[Sch\"uhle \etal\ (2000)]{Schuehle2000}
{Sch{\"u}hle, U., Wilhelm, K., Hollandt, J., Lemaire, P., \& Pauluhn, A.} 2000, \textit{Astronomy and Astrophysics}, 354, L71

\bibitem[Shapiro \etal\ (2009)]{Shapiro2009}
{Shapiro, A.I., Schmutz, W., Sch\"oll, M., Haberreiter, M., \& Rozanov, E.} 2009, \textit{Astronomy and Astrophysics}, to be submitted

\bibitem[Solanki \etal\ (2005)]{Solanki2005}
{Solanki, S.K., Krivova, N.A., \& Wenzler, T.} 2005, \textit{Advances in Space Research}, 35, 376

\bibitem[Steinhilber \etal\ (2009)]{Steinhilber2009}
{Steinhilber, F., Beer, J., \& {Fr{\"o}hlich}, C.} 2009, \textit{Geophysical Research Letters}, 36, 19704
 
\bibitem[Thuillier \etal\ (2003)]{Thuillier2003}
{Thuillier, G., Hers{\' e}, M., Labs, D., Foujols, T., Peetermans, W., Gillotay, D.,  Simon, P.C., \& Mandel, H.} 2003, \textit{Solar Physics}, 214, 1


\bibitem[White \etal\ (2000)]{White2000SSRv}
{White, O.R., Fox, P.A., Meisner, R., Rast, M.P., Yasukawa, E., Koon, D., Rice, C., Lin, H., Kuhn, J., \& Coulter, R.} 2000, \textit{Space Science Reviews}, 94, 75

\bibitem[Willson (1994)]{Willson1994}
{Willson, B.C.} 1994, in: J.M. Pap, C. Frolich, H.S. Hudson, \& S. Solanki (eds.), \textit{Proceedings from IAU Colloquium 143: The Sun as a Variable Star: Solar and Stellar Irradiance Variations}, p. 54

\bibitem[Woods \etal\ (2000)]{Woods2000}
{Woods, T.N., Tobiska, W.K., Rottman, G.J., Worden, J.R.} 2000, \textit{Journal of Geophysical Research}, 105, 27195

\bibitem[Woods \etal\ (2006)]{Woods2006}
{Woods, T.N., Lean, J.L., Eparvier, F.G.} 2006, in: N. Gopalswamy, \& A. {Bhattacharyya} (eds.), \textit{Proceedings of the ILWS Workshop}, p.145

\end{thebibliography}
\end{document}